\documentclass{aastex631}
\usepackage{natbib}

\begin{document}

\title{Investigating the Correlation between Dark Matter Content, Ages and Mass-to-Light Ratios in Spiral Galaxies}

\author{Arpit Kottur}
\affiliation{Department of Physics \\
Fergusson College (Autonomous) \\
Shivajinagar, Pune, India}

\author{Meet Mehta}
\affiliation{Department of Physics \\
Fergusson College (Autonomous) \\
Shivajinagar, Pune, India}

\author{Raka Dabhade}
\affiliation{Department of Physics \\
Fergusson College (Autonomous) \\
Shivajinagar, Pune, India}

%\collaboration{20}{(AAS Journals Data Editors)}

%\author{F.X Timmes}
%\affiliation{Arizona State University}
%\affiliation{AAS Journals Associate Editor-in-Chief}

%\altaffiliation{AASTeX v6+ programmer}

%\affiliation{AAS Director of Publishing}
%\affiliation{American Astronomical Society \\
%1667 K Street NW, Suite 800 \\



\begin{abstract}

We present an empirical investigation into the relationship between galactic age and dark matter content across a sample of 16 nearby, well-resolved spiral galaxies. Using raw rotation curve data from IOA Tokyo's publicly available repository, we model each galaxy's mass distribution via a three-component decomposition—Hernquist bulge, exponential disk, and a Navarro-Frenk-White (NFW) dark matter halo—fit using Monte Carlo simulations. The onset of dark matter dominance was identified using the NFW scale radius, beyond which we computed the total enclosed mass via Keplerian dynamics. I-band luminosities for these regions were estimated using a calibrated Tully-Fisher relation, yielding precise mass-to-light (M/L) ratios. We further calculated dark matter mass and density using NFW profile equations, and galaxy ages were retrieved through an extensive literature survey of stellar population studies.

Our analysis reveals strong positive correlations between galactic age and both dark matter mass (Pearson $r \approx 0.91$; Spearman $\rho \approx 0.93$) and density (Pearson $r \approx 0.91$; Spearman $\rho \approx 0.91$), as well as M/L ratios, suggesting a robust link between evolutionary history and dark matter build-up. These findings are in quantitative agreement with predictions from large-scale cosmological simulations that incorporate \textit{assembly bias} and \textit{smooth accretion-dominated growth}, reinforcing the view that older galaxies, having formed earlier in high-density peaks, have accumulated significantly more dark matter over cosmic time. Our results offer observational evidence for time-dependent dark matter assembly and establish galactic age as a meaningful tracer of halo evolution.
\end{abstract}



\section{Introduction} \label{sec:intro}

The flat rotation curves observed in spiral galaxies have posed a longstanding enigma in astrophysics, providing compelling evidence for the existence of dark matter. While the nature of this elusive component remains a subject of intense investigation, understanding the distribution of baryonic matter within galaxies and its relationship to their dynamical properties is essential for unraveling the complexities of galaxy formation and evolution.\\

This study explores the intricate connections between galactic rotation curves (GRCs), mass-to-luminosity ratios (M/L), and the probable age estimates of spiral galaxies. By examining a sample of 20 galaxies, we aim to elucidate the interplay between these parameters and their implications for galactic structure and evolution.

\subsection{Galactic Rotation Curves}\label{subsec: GRC}

A galactic rotation curve is a graphical representation of the orbital velocities of stars and gas within a galaxy as a function of their distance from the galactic center. These curves are typically derived by averaging data from both sides of a galaxy, as observed velocities often exhibit asymmetries. A fundamental discrepancy arises when comparing these observed rotation curves to theoretical predictions based on the visible matter content of the galaxy and Newtonian gravity.\\ 

Unlike planetary systems where orbital velocities decrease with distance from the central body according to Kepler's Third Law, the orbital velocities of stars within galaxies tend to remain constant or even increase at larger radii. This unexpected behavior indicates a mass distribution that is far more extended than can be accounted for by the visible matter alone. The resulting mismatch between the observed mass distribution and the mass inferred from rotation curves has been a cornerstone in the development of the dark matter hypothesis. (\cite{rubin1978extended} , \cite{rubin1983rotation}), (\cite{sofue2001rotation})

\subsection{Mass-to-Light Ratios}\label{subsec: ML Ratios}

A fundamental parameter in characterizing galaxies is the mass-to-light ratio (M/L), which represents the ratio of a galaxy's mass to its luminosity. A galaxy composed entirely of stars identical to the Sun would exhibit an M/L ratio of 1. However, the diverse stellar populations within galaxies, encompassing stars of varying ages and masses, result in a more complex picture.\\

Generally, galaxies exhibit M/L ratios ranging from 1 to 30. Irregular galaxies, dominated by young, luminous stars, tend to have lower M/L values, while elliptical galaxies, with predominantly old, less luminous stellar populations, exhibit higher M/L ratios. Spiral galaxies, with a mix of stellar populations, occupy an intermediate range. \\

The underlying reason for these variations lies in the relationship between stellar mass and luminosity. High-mass stars, though rare, are exceptionally luminous, contributing disproportionately to a galaxy's overall brightness while representing a smaller fraction of its total mass. Conversely, low-mass stars, while numerous, contribute relatively little to a galaxy's luminosity but constitute a significant portion of its mass. Furthermore, the evolution of stellar populations influences M/L ratios. As massive stars evolve off the main sequence and become compact white dwarfs, the overall M/L of a galaxy increases.\\

It is crucial to note that mass-to-light ratios derived from visible light only account for the galaxy's stellar component. The discovery of dark matter has necessitated a reevaluation of these calculations. While stellar populations impose an upper limit on M/L values, ratios significantly exceeding this threshold provide compelling evidence for the existence of non-baryonic matter within galaxies. (\cite{schwarzschild1954mass}), (\cite{takamiya2000radial})

\subsection{Probable Age Estimates}\label{subsec: Age}

Accurately determining the age of a galaxy presents a significant challenge in astrophysics. While direct measurement is not feasible, astronomers employ sophisticated techniques to estimate galactic ages based on the properties of their stellar populations. One such method involves stellar population synthesis, wherein a galaxy's observed spectrum is compared to theoretical models of stellar populations with varying ages and metallicities. By identifying the optimal match, researchers can infer the galaxy's approximate age. \\

Additionally, the color-magnitude diagram (CMD) of a galaxy's stars offers valuable clues. Older galaxies tend to exhibit redder colors due to a preponderance of low-mass, long-lived stars, while younger galaxies are characterized by bluer colors associated with a higher proportion of massive, short-lived stars. \\

It is essential to acknowledge that uncertainties inherent in stellar evolution models and observational limitations introduce challenges in precise age determination. Nonetheless, these techniques provide essential insights into the formation and evolutionary history of galaxies. (\cite{thomas2017extended})

\section{Data and Methodology} \label{sec: Data and Methodology}

\subsection{Data Sources} \label{subsec: Sources}

The rotation curve data for this study were obtained from the publicly accessible repository maintained by Dr. Yoshiaki Sofue at the Institute of Astronomy, University of Tokyo. This comprehensive dataset compiles high-resolution rotation curves for nearby spiral galaxies, constructed from combined observations in the CO, HI, and optical regimes. (\cite{sofue1997nuclear}), (\cite{sofue1999central}), (\cite{takamiya2002iteration}), (\cite{sofue2016rotation})

Our analysis focuses on a sample of 16 galaxies selected based on the following criteria:
\begin{itemize}
    \item High spatial resolution and radial extent of the rotation curve;
    \item Availability of well-separated velocity contributions from bulge, disk, and halo components;
    \item Independent age estimates available in the literature;
    \item Minimal inclination uncertainty and kinematic asymmetries.
\end{itemize}

The original data include tabulated values of radial distance (in kiloparsecs) and corresponding rotational velocity (in km s$^{-1}$), along with individual decomposed components for the stellar bulge, stellar disk, and the dark matter halo based on standard mass modeling. These decompositions are derived using a parametric approach outlined in \cite{sofue2016rotation}, enabling a component-wise reconstruction of each galaxy’s total rotation curve.

Our initial candidate list consisted of 20 spiral galaxies selected from (\cite{sofue2016rotation}) rotation curve catalog. To ensure the robustness of our correlation analysis, we applied strict exclusion criteria regarding the availability of independent age estimates. Four galaxies (NGC 598, NGC 1068, NGC 1417 and NGC 3521) were excluded from the final sample because their stellar population age estimates in the literature were either highly unconstrained or unavailable. The final sample of 16 galaxies represents the subset for which both high-resolution kinematic data and robust, independent evolutionary timelines are available.

We use this dataset as the foundation for all subsequent mass modeling, Monte Carlo fitting, and dark matter profile analysis described in the following subsections.

\subsection{Mass Modeling and Monte Carlo Fitting} \label{subsec: Mass and Monte Carlo}

To reconstruct the full mass distribution in each galaxy, we adopted a three-component model comprising a stellar bulge, a stellar disk, and a dark matter halo. The bulge was modeled using the Hernquist profile, the disk as an exponential thin disk, and the halo with the Navarro–Frenk–White (NFW) density profile. (\cite{hernquist1990analytical}), (\cite{freeman1970disks}), (\cite{navarro1997universal}), (\cite{binney2011galactic}), (\cite{klypin2001resolving})

The total circular velocity at any radial distance $r$ is given by the quadratic sum of the individual velocity components:
\begin{equation}
    V_{\text{tot}}^2(r) = V_{\text{bulge}}^2(r) + V_{\text{disk}}^2(r) + V_{\text{halo}}^2(r),
\end{equation}
where each component is defined by the respective mass profile and structural parameters. The Hernquist bulge component is expressed as:
\begin{equation}
    V_{\text{bulge}}(r) = \sqrt{ \frac{G M_b r^2}{(r + a_b)^2} },
\end{equation}
with $M_b$ the bulge mass and $a_b$ the scale radius. The disk velocity follows from the exponential disk potential:
\begin{equation}
    V_{\text{disk}}(r) = \sqrt{ \frac{G M_d}{2 R_d} y^2 \left[ I_0(y) K_0(y) - I_1(y) K_1(y) \right] },
\end{equation}
where $y = r / (2R_d)$, and $I_n$, $K_n$ are modified Bessel functions of the first and second kinds, respectively. $M_d$ is the disk mass, and $R_d$ is the disk scale length.

The NFW halo component is given by:
\begin{equation}
    V_{\text{halo}}(r) = \sqrt{ 4\pi G \rho_s r_s^3 \left[ \frac{\ln(1 + r/r_s) - \frac{r/r_s}{1 + r/r_s}}{r} \right] },
\end{equation}
where $\rho_s$ is the characteristic halo density, and $r_s$ is the scale radius.

To determine the best-fit parameters for each component, we implemented a Monte Carlo fitting algorithm. For each galaxy, we generated $10^4$ realizations by randomly sampling the parameter space within physically motivated priors. The priors for bulge mass, disk scale length, and halo concentration were selected based on typical spiral galaxy values drawn from \cite{sofue2016rotation} and \cite{bullock2001profiles}.

Each realization produced a synthetic rotation curve, which was compared to the observed data using a chi-squared minimization criterion:
\begin{equation}
    \chi^2 = \sum_{i} \frac{\left[ V_{\text{obs}}(r_i) - V_{\text{model}}(r_i) \right]^2}{\sigma_i^2},
\end{equation}
where $V_{\text{obs}}(r_i)$ is the observed velocity at radius $r_i$, $V_{\text{model}}(r_i)$ is the model velocity, and $\sigma_i$ is the observational uncertainty. The best-fit parameters were taken as the median values from the posterior distributions of all realizations with $\chi^2$ values within the 1$\sigma$ confidence level of the global minimum.

We made use of the RotCurveTool, a Rotation Curve Fitting tool designed to fit our specific requirements (\cite{kottur2025rotcurvetool}). This approach ensured both robustness and proper treatment of degeneracies in parameter space, and it enabled us to extract precise estimates for each galaxy’s structural parameters, especially the NFW scale radius $r_s$, which plays a key role in identifying the dark matter-dominated region in subsequent analyses.

\begin{figure}[h]
    \centering
    \includegraphics[width=1\linewidth]{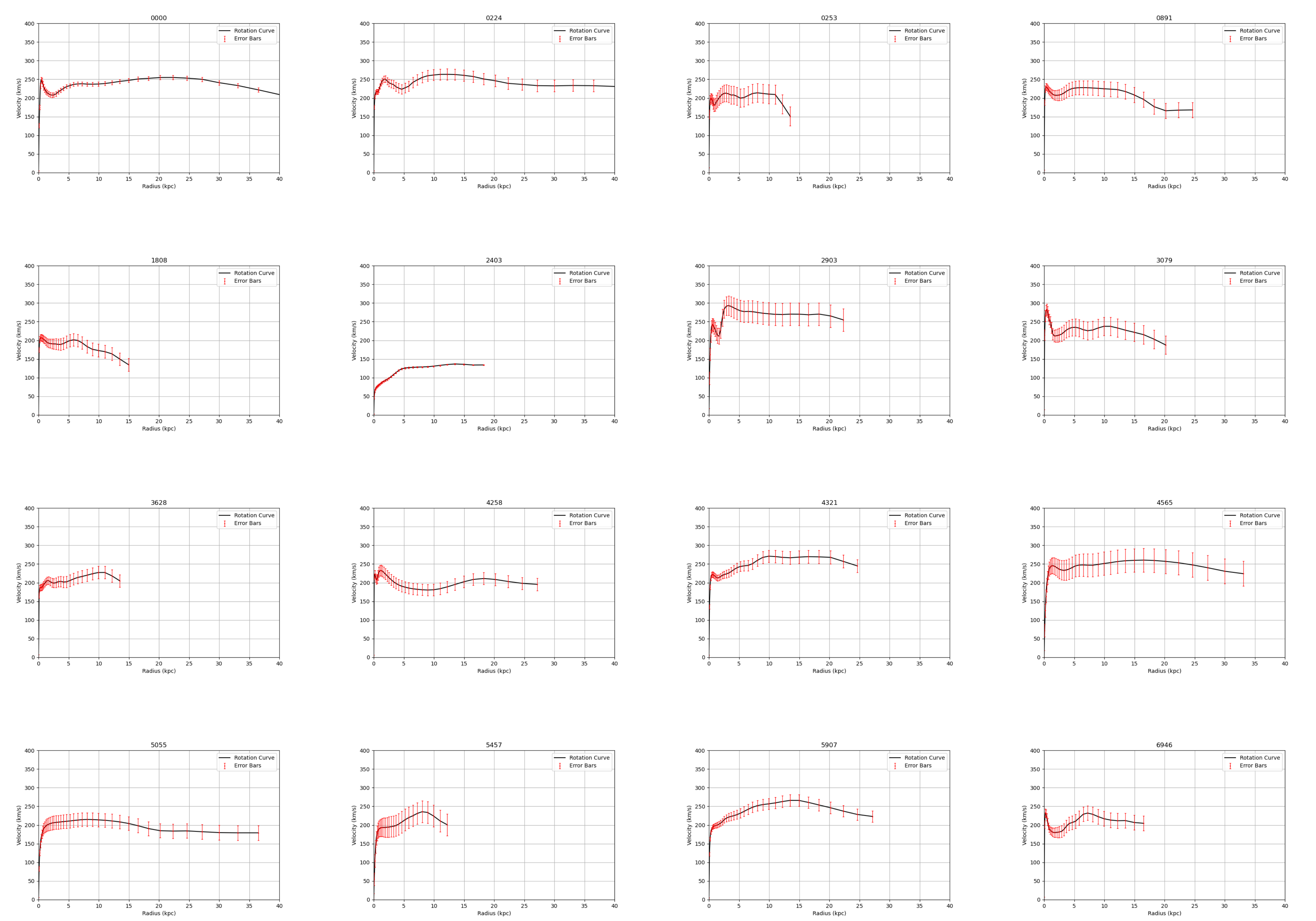}
    \caption{\label{GRC Catalogue}Decomposed rotation curves for the sample of 16 spiral galaxies. The black points with error bars represent the observed raw rotation velocities derived from combined CO, HI, and optical observations.}
\end{figure}

\subsection{Identification of the Dark Matter Dominated Region and Mass Calculation} \label{subsec: DM Dominated Region}

To isolate the region within each galaxy where dark matter dominates the gravitational potential, we adopted the scale radius $r_s$ derived from the best-fit Navarro–Frenk–White (NFW) halo profile (\cite{navarro1997universal}). The scale radius corresponds to the point at which the logarithmic slope of the NFW density profile equals $-3$, marking a transition in the halo’s structural behavior. Beyond this radius, the dark matter halo's contribution to the total circular velocity begins to surpass that of the baryonic components in most spiral galaxies.(\cite{sofue2016rotation}

For each galaxy, we defined the dark matter-dominated region as the radial interval $r \geq r_s$. This conservative threshold ensures minimal contamination from baryonic mass while maintaining physical relevance for outer halo dynamics.

The total mass enclosed within this DM-dominated region, $M_{\text{DM}}(r \geq r_s)$, was estimated using the Keplerian approximation:
\begin{equation}
    M_{\text{enc}}(r) = \frac{r \cdot V_{\text{tot}}^2(r)}{G},
\end{equation}
where $V_{\text{tot}}(r)$ is the total circular velocity at radius $r$, and $G$ is the gravitational constant. The value of $r$ was chosen slightly beyond $r_s$, corresponding to  $1\sigma$, to avoid numerical instabilities and to account for measurement uncertainties.

To maintain consistency with observational data and rotation curve resolution, we propagated the uncertainties in both $V_{\text{tot}}$ and $r$ using standard error propagation techniques (\cite{andrae2010error}). The resulting enclosed mass, $M_{\text{enc}}(r \geq r_s)$, was interpreted as the total mass predominantly composed of dark matter, under the assumption that baryonic contributions become subdominant in this regime (\cite{navarro1997universal}.

These mass estimates form the basis for our calculation of mass-to-light ratios and dark matter densities in the subsequent steps of the analysis.

\begin{figure}[h]
    \centering
    \includegraphics[width=1\linewidth]{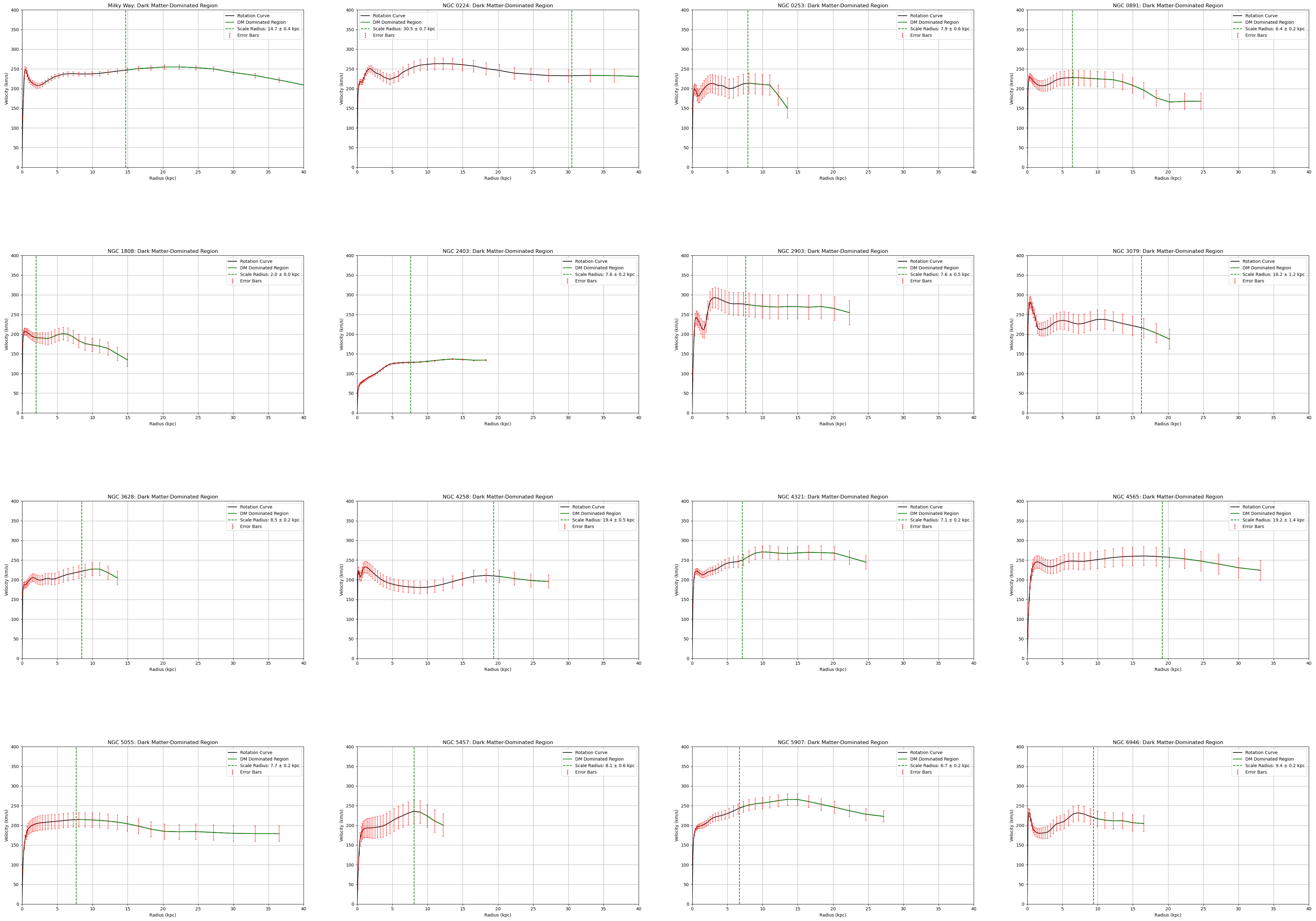}
    \caption{\label{GRC Catalogue-DM}Identification of Dark Matter dominated regions. The vertical line marks the NFW scale radius ($r_s$) for each galaxy. The region to the right ($r \ge r_s$) is defined as the dark matter-dominated envelope. The total enclosed mass ($M_{enc}$) and I-band luminosity were calculated specifically within this outer radial domain to minimize baryonic contamination.}
\end{figure}

\textbf{Note:} The data used for analysis extends till the specific $R_{200}$ values of each galaxy obtained from \cite{sofue2016rotation} and NFW modeling , as opposed to the 40 kpc plotted from the raw data.

\subsection{Luminosity Estimation Using the Tully--Fisher Relation} \label{subsec: Luminosty using TFR}
A key methodological challenge in this study is estimating the baryonic luminosity specifically within the dark-matter-dominated outer envelope ($r > r_s$). Standard photometric catalogs (e.g., NED, SIMBAD) typically provide integrated magnitudes for the entire galaxy or within fixed isophotal apertures, which are heavily dominated by the inner disk and bulge components. Spatially resolved surface photometry profiles extending into the faint outer halo are not uniformly available for our heterogeneous sample.

To overcome this, we employ the Tully-Fisher relation,  which provides a well-calibrated empirical link between a galaxy’s intrinsic luminosity and its rotational velocity, (\cite{tully1977new}), (\cite{tully2000distances}), (\cite{masters2006sfi++}), (\cite{reyes2011calibrated}) to estimate the characteristic luminosity associated with the halo's virial dynamics. While we acknowledge that this introduces a dependence on rotational velocity, it provides a physically motivated proxy for the expected luminosity of the system's potential well, essentially predicting the 'missing light' that corresponds to the observed halo mass. To mitigate circularity concerns, we treat this derived luminosity strictly as a tracer of the system's total baryonic capacity rather than a direct photometric observation

The Tully–Fisher relation in the I-band takes the general form:
\begin{equation}
    M_I = a \log_{10}(V_{\text{rot}}) + b,
\end{equation}
where $M_I$ is the absolute I-band magnitude, $V_{\text{rot}}$ is the characteristic rotation velocity (typically taken as the flat part of the rotation curve or the maximum value), and $a$, $b$ are calibration constants. We adopted the parameter values from \cite{masters2006sfi++} and \cite{reyes2011calibrated}, which are based on large galaxy samples and provide reliable fits across various morphological types.

For consistency with our dark matter region definition, we used $V_{\text{rot}} = V_{\text{tot}}(r = 1.1\,r_s)$, i.e., the rotation velocity just beyond the NFW scale radius, as a proxy for the flat rotational velocity.

The absolute magnitude was converted to I-band luminosity using:
\begin{equation}
    L_I = L_{I,\odot} \times 10^{-0.4(M_I - M_{I,\odot})},
\end{equation}
where $M_{I,\odot} = 4.15$ is the absolute I-band magnitude of the Sun, and $L_{I,\odot}$ is the solar I-band luminosity in physical units. This yields the luminosity associated with the dark matter-dominated portion of the galaxy.

Uncertainties in $V_{\text{rot}}$ were propagated through both equations using standard error analysis \cite{andrae2010error}, and systematic errors arising from TFR scatter were incorporated as an additional $\sim 0.1$ mag dispersion in $M_I$ following \cite{reyes2011calibrated}.

This luminosity estimate was then used to compute the mass-to-light ratio in the dark matter-dominated region.

\subsection{Calculation of Mass-to-Light Ratio, Dark Matter Mass, and Density}\label{subsec: Calculations}

With both the total enclosed mass $M_{\text{enc}}$ and the I-band luminosity $L_I$ estimated for the dark matter-dominated region ($r \geq r_s$), we computed the mass-to-light ratio (M/L) as:
\begin{equation}
    \left( \frac{M}{L} \right)_I = \frac{M_{\text{enc}}(r \geq r_s)}{L_I},
\end{equation}
where both mass and luminosity correspond to the same radial domain. This ratio provides insight into the relative dominance of dark matter in regions where baryonic matter is dynamically subdominant.

Additionally, we calculated the dark matter mass and density directly from the best-fit NFW halo parameters. The enclosed dark matter mass within a radius $r$ in the NFW profile is given by:
\begin{equation}
    M_{\text{DM}}(r) = 4\pi \rho_s r_s^3 \left[ \ln \left(1 + \frac{r}{r_s} \right) - \frac{r}{r + r_s} \right],
\end{equation}
where $\rho_s$ and $r_s$ are the characteristic density and scale radius of the NFW halo, respectively \cite{navarro1997universal}.

We evaluated $M_{\text{DM}}$ at $r = 1.1\,r_s$, consistent with our earlier mass and velocity estimates. The corresponding average dark matter density over the same region was computed as:
\begin{equation}
    \rho_{\text{DM}} = \frac{M_{\text{DM}}(r)}{\frac{4}{3}\pi r^3}.
\end{equation}

Uncertainties in $M_{\text{enc}}$, $L_I$, and NFW parameters were propagated using Monte Carlo resampling across $10^4$ realizations per galaxy. (\cite{andrae2010error})

This final set of measurements formed the basis for the correlation analysis, where we investigate trends with galactic age and compare them with theoretical expectations from cosmological simulations.

\subsection{Probable Age Estimates}\label{subsec: Age Estimates}

To investigate correlations between dark matter content and galactic age, we compiled stellar population age estimates for each of the 16 galaxies in our sample from a comprehensive literature survey. These ages were drawn from a range of studies employing different methodologies, including color–magnitude diagram (CMD) fitting, integrated spectral synthesis, resolved stellar population modeling, and stellar kinematic analysis.

Whenever available, we prioritized studies with:
\begin{itemize}
    \item Spatially resolved star formation histories,
    \item Multiband photometry or spectroscopic age diagnostics,
    \item Explicit age estimates for bulge, disk, or halo components, preferably reported in gigayears (Gyr),
    \item Uncertainties or age ranges derived from statistical modeling.
\end{itemize}

In cases where multiple sources provided consistent results, we adopted the median value. For galaxies with component-specific ages (e.g., disk vs. bulge), we used luminosity-weighted averages unless a single dominant population was clearly identified.

The references used span a variety of high-quality observational programs, including:
\begin{itemize}
\item \cite{mouhcine2007stellar}, 
\item \cite{greggio2014panoramic}, 
\item \cite{vermot20233d}, 
\item \cite{xiang2022time}, 
\item \cite{leahy2023andromeda}, 
\item \cite{bell2000stellar}, 
\item \cite{laine2016metallicity}, 
\item \cite{kormendy2019structural}, 
\item \cite{garner2024dynamic}, 
\item \cite{kostiuk2025comprehensive}, 
\item \cite{pessa2023resolved}, 
\item \cite{lofaro2024ancient}, 
\item \cite{duan2006multicolor}, 
\item \cite{thomas2011dynamical}, 
\item \cite{carrillo2020virus}, 
\item \cite{kang2017role}, 
\item \cite{scott2017sami}, 
\item \cite{terlevich2002catalogue}, 
\item \cite{da1984age}, 
\item \cite{tacconi1996sub}, 
\item \cite{monkiewicz1999age}, 
\item \cite{thomas2017extended}, 
\item \cite{williams2013ages}.
\end{itemize}

We acknowledge that compiling age estimates from heterogeneous sources introduces systematic uncertainties due to differing Initial Mass Functions (IMFs), stellar population synthesis models, and dust extinction corrections. However, our primary focus is on the relative evolutionary stage of these systems rather than absolute chronology. To mitigate the impact of model-dependent systematics, we employ the Spearman rank correlation coefficient ($\rho$) alongside the Pearson coefficient. The Spearman test relies solely on the monotonic ranking of galaxies by age, making it robust against systematic shifts in absolute age calibration.

These age estimates serve as the independent variable in our statistical tests correlating dark matter mass, density, and M/L ratios with galactic age.

\subsection{Correlation Analysis}\label{subsec: Correlation Analysis}

To quantify the relationships between galactic age and various derived dark matter properties, we performed both parametric and non-parametric correlation tests. Specifically, we computed: (\cite{artusi2002bravais})

\begin{itemize}
    \item The \textbf{Pearson correlation coefficient} ($r$), which measures the degree of linear correlation between two continuous variables $(x, y)$. It is defined as:
    \begin{equation}
        r = \frac{\sum_{i=1}^{n}(x_i - \bar{x})(y_i - \bar{y})}{\sqrt{\sum_{i=1}^{n}(x_i - \bar{x})^2 \sum_{i=1}^{n}(y_i - \bar{y})^2}},
    \end{equation}
    where $\bar{x}$ and $\bar{y}$ are the sample means of $x$ and $y$, respectively. The value of $r$ lies in the range $[-1, 1]$, with $r = 1$ indicating perfect positive linear correlation.

    \item The \textbf{Spearman rank correlation coefficient} ($\rho$), which assesses the strength of a monotonic relationship by computing the Pearson correlation on rank-transformed data. It is given by:
    \begin{equation}
        \rho = 1 - \frac{6 \sum_{i=1}^{n} d_i^2}{n(n^2 - 1)},
    \end{equation}
    where $d_i$ is the difference in rank between $x_i$ and $y_i$, and $n$ is the sample size.
\end{itemize}

To assess the statistical significance of the correlations, we performed two-tailed hypothesis tests under the null hypothesis $H_0$: no correlation exists between the variables. The associated $p$-values were computed analytically for Pearson $r$, and via permutation-based tests for Spearman $\rho$.

\begin{table*}[h]
\centering
\caption{Derived parameters for the galaxy sample. This table lists the I-band Mass-to-Light ratio, independent probable age estimates derived from a comprehensive literature survey of stellar populations, the total enclosed Dark Matter Mass ($M_{DM}$) within the halo-dominated region, and the characteristic Dark Matter Density ($\rho_{DM}$). Uncertainties were calculated using Monte Carlo error propagation.}
\label{tab:galaxy_parameters}
\begin{tabular}{ccccccccc}
\hline
\textbf{NGC} & $(M/L)_I$ & $\Delta(M/L)_I$ & Age (Gyr) & $\Delta$Age (Gyr) & $M_{\rm DM}$ ($M_{\odot}$) & $\Delta M_{\rm DM}$ ($M_{\odot}$) & $\rho_{\rm DM}$ ($\frac{M_{\odot}}{\text{kpc}^3}$) & $\Delta \rho_{\rm DM}$ \\
\hline
0000 & 8.052 & 2.369 & 13.0 & 0.5 & $7.18 \times 10^{11}$ & $1.06 \times 10^{11}$ & $2.50 \times 10^{4}$ & $5.70 \times 10^{3}$ \\
0224 & 12.188 & 3.032 & 12.0 & 1.5 & $8.65 \times 10^{11}$ & $2.10 \times 10^{10}$ & $2.44 \times 10^{4}$ & $4.30 \times 10^{3}$ \\
0253 & 4.453 & 2.251 & 8.5 & 2.5 & $4.25 \times 10^{11}$ & $1.35 \times 10^{11}$ & $1.55 \times 10^{4}$ & $6.10 \times 10^{3}$ \\
0891 & 5.528 & 2.961 & 10.0 & 1.0 & $5.19 \times 10^{11}$ & $5.70 \times 10^{10}$ & $1.73 \times 10^{4}$ & $7.40 \times 10^{3}$ \\
1808 & 3.200 & 2.024 & 7.0 & 2.5 & $8.57 \times 10^{10}$ & $9.50 \times 10^{9}$ & $9.10 \times 10^{3}$ & $4.00 \times 10^{2}$ \\
2403 & 8.043 & 2.128 & 6.8 & 1.5 & $2.62 \times 10^{11}$ & $3.50 \times 10^{10}$ & $1.54 \times 10^{4}$ & $4.80 \times 10^{3}$ \\
2903 & 4.258 & 2.202 & 7.0 & 2.0 & $3.10 \times 10^{11}$ & $2.90 \times 10^{10}$ & $1.57 \times 10^{4}$ & $8.40 \times 10^{3}$ \\
3079 & 7.500 & 3.443 & 9.0 & 2.0 & $7.04 \times 10^{11}$ & $1.96 \times 10^{11}$ & $2.19 \times 10^{4}$ & $1.58 \times 10^{4}$ \\
3628 & 4.125 & 1.282 & 6.5 & 0.05 & $2.72 \times 10^{11}$ & $6.90 \times 10^{10}$ & $1.29 \times 10^{4}$ & $7.30 \times 10^{3}$ \\
4258 & 9.652 & 3.096 & 12.0 & 1.0 & $9.18 \times 10^{11}$ & $1.70 \times 10^{10}$ & $2.47 \times 10^{4}$ & $5.60 \times 10^{3}$ \\
4321 & 4.423 & 1.933 & 8.0 & 2.0 & $3.18 \times 10^{11}$ & $7.10 \times 10^{10}$ & $1.40 \times 10^{4}$ & $7.70 \times 10^{3}$ \\
4565 & 8.793 & 3.628 & 12.0 & 1.0 & $8.51 \times 10^{11}$ & $5.00 \times 10^{10}$ & $2.50 \times 10^{4}$ & $1.55 \times 10^{4}$ \\
5055 & 8.075 & 4.611 & 11.0 & 1.0 & $8.09 \times 10^{11}$ & $5.90 \times 10^{10}$ & $2.27 \times 10^{4}$ & $5.80 \times 10^{3}$ \\
5457 & 3.973 & 1.994 & 7.0 & 2.0 & $3.54 \times 10^{11}$ & $8.90 \times 10^{10}$ & $1.50 \times 10^{4}$ & $7.30 \times 10^{3}$ \\
5907 & 4.967 & 2.261 & 8.0 & 1.5 & $3.55 \times 10^{11}$ & $5.30 \times 10^{10}$ & $1.62 \times 10^{4}$ & $5.10 \times 10^{3}$ \\
6946 & 5.094 & 1.998 & 8.5 & 1.0 & $4.24 \times 10^{11}$ & $7.40 \times 10^{10}$ & $1.59 \times 10^{4}$ & $7.00 \times 10^{3}$ \\
\hline
\end{tabular}
\end{table*}

This procedure ensures a statistically robust and uncertainty-aware assessment of the correlation strengths. The full results and interpretation are provided in Section \ref{sec: Results}.

\section{Results}\label{sec: Results}
We present the results of our statistical correlation analysis between key galaxy parameters derived in Section~2. This includes the total dark matter mass ($M_{\rm DM}$), average halo density ($\rho_{\rm DM}$), mass-to-light ratio ($(M/L)_I$), and galactic age. For each pairwise relationship, we compute the Pearson ($r$) and Spearman ($\rho$) correlation coefficients based on $10^4$ Monte Carlo realizations, along with their associated $p$-values. 

\begin{table}[h]
\centering
\caption{Summary of statistical correlations. The table presents the Pearson ($r$) and Spearman ($\rho$) correlation coefficients for pairwise comparisons of galactic age, dark matter mass ($M_{DM}$), halo density ($\rho_{DM}$), and mass-to-light ratios ($(M/L)_I$). The associated p-values ($<0.001$) indicate that the strong positive correlations observed are statistically significant.}
\label{tab:corr_summary}
\begin{tabular}{lccc}
\hline
\textbf{Parameter Pair} & \textbf{Pearson $r$} & \textbf{Spearman $\rho$} & \textbf{$p$-value} \\
\hline
$M_{\rm DM}$ vs Age              & 0.913 & 0.935 & $< 0.001$ \\
$\rho_{\rm DM}$ vs Age           & 0.912 & 0.913 & $< 0.001$ \\
$M_{\rm DM}$ vs $(M/L)_I$        & 0.847 & 0.838 & $< 0.001$ \\
$\rho_{\rm DM}$ vs $(M/L)_I$     & 0.866 & 0.874 & $< 0.001$ \\
Age vs $(M/L)_I$                 & 0.782 & 0.803 & $< 0.001$ \\
\hline
\end{tabular}
\end{table}

Our statistical analysis reveals a coherent set of correlations linking galactic evolutionary history with dark matter halo properties. As summarized in Table \ref{tab:corr_summary}, we find strong, positive monotonic relationships across all tested parameters.

\begin{figure}[h!]
    \centering
    \includegraphics[width=0.5\linewidth]{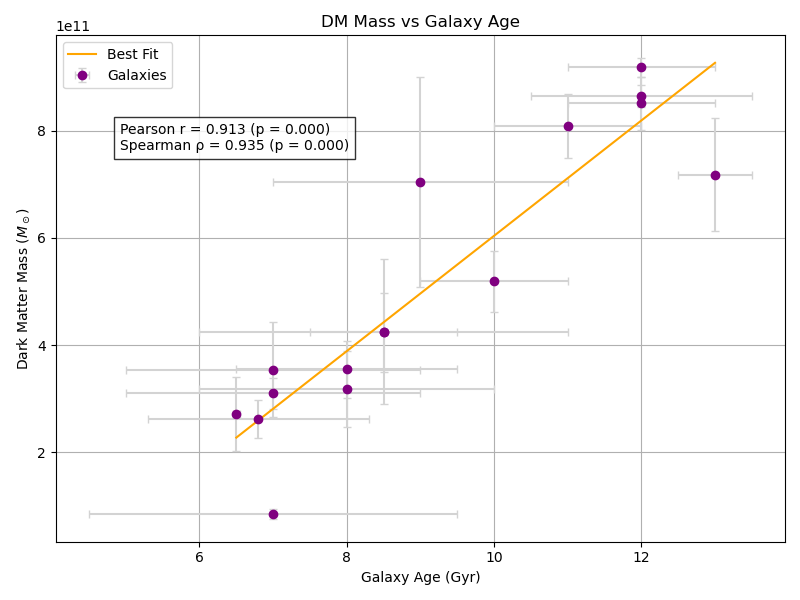}
    \caption{\label{DMvAge}Correlation between Dark Matter Mass ($M_{DM}$) and Galactic Age. The scatter plot shows a strong positive correlation (Pearson $r \approx 0.91$) between the age of the stellar population and the mass of the dark matter halo. The solid line represents the linear best fit. This trend supports the ``smooth accretion'' scenario, suggesting that older galaxies have had more cosmic time to accumulate diffuse dark matter.}
\end{figure}

\begin{figure}[h!]
    \centering
    \includegraphics[width=0.5\linewidth]{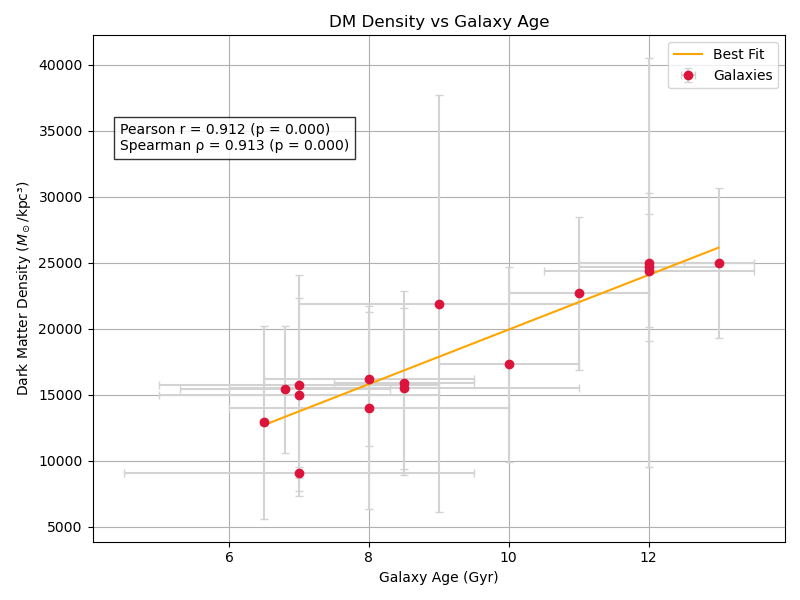}
    \caption{\label{DMdvAge}Correlation between Dark Matter Density ($\rho_{DM}$) and Galactic Age. The data reveals that older galaxies possess significantly denser dark matter halos (Pearson $r \approx 0.91$). This result is quantitatively consistent with cosmological ``assembly bias'' simulations, which predict that early-forming halos are more concentrated than those forming later.}
\end{figure}

First, we observe a distinct dependence of dark matter content on galactic age. As shown in Figure \ref{DMvAge} and \ref{DMdvAge}, older galaxies host systematically larger dark matter masses ($ r \approx 0.91$) and exhibit higher characteristic halo densities ($r \approx 0.91$). This suggests that halo assembly is a time-cumulative process, where older systems have had more time to accrete mass and virialise into denser configurations.

\begin{figure}[h!]
    \centering
    \includegraphics[width=0.5\linewidth]{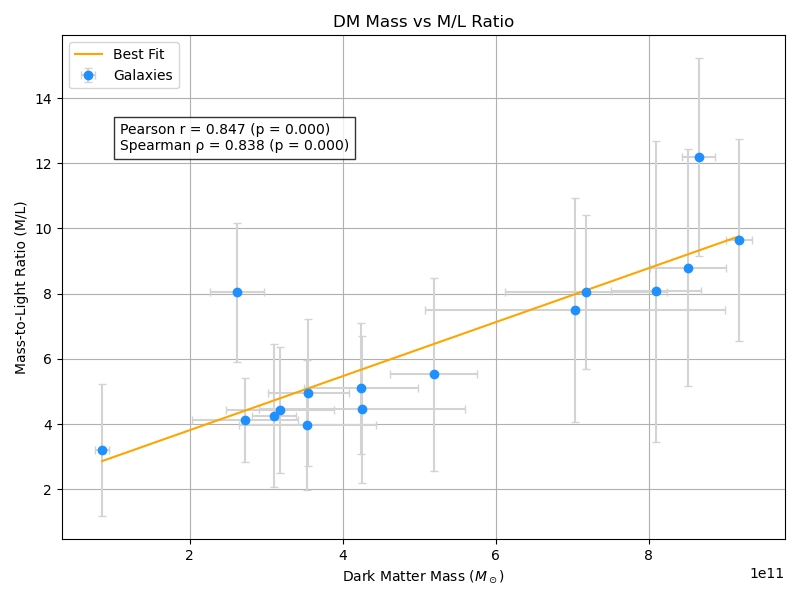}
    \caption{\label{DMvML}Dark Matter Mass ($M_{DM}$) vs. Mass-to-Light Ratio ($(M/L)_I$). The positive correlation (Pearson $r \approx 0.85$) indicates that galaxies with more massive halos are increasingly dark matter-dominated relative to their optical luminosity.}
\end{figure}

\begin{figure}[h!]
    \centering
    \includegraphics[width=0.5\linewidth]{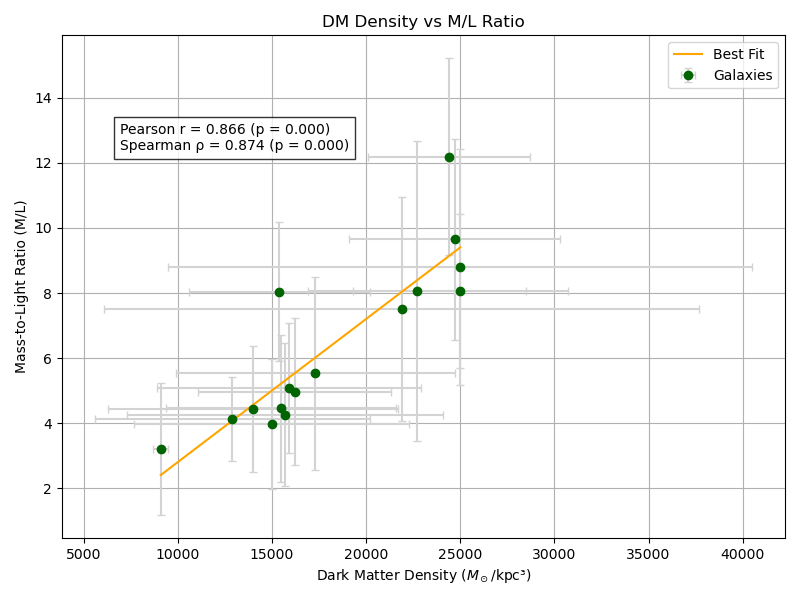}
    \caption{\label{DMdvML}Dark Matter Density ($\rho_{DM}$) vs. Mass-to-Light Ratio ($(M/L)_I$). A strong monotonic relationship (Spearman $\rho \approx 0.87$) connects the compactness of the halo to the galaxy's mass-to-light ratio, further linking dynamical structure to baryonic efficiency.}
\end{figure}

\begin{figure}[h!]
    \centering
    \includegraphics[width=0.5\linewidth]{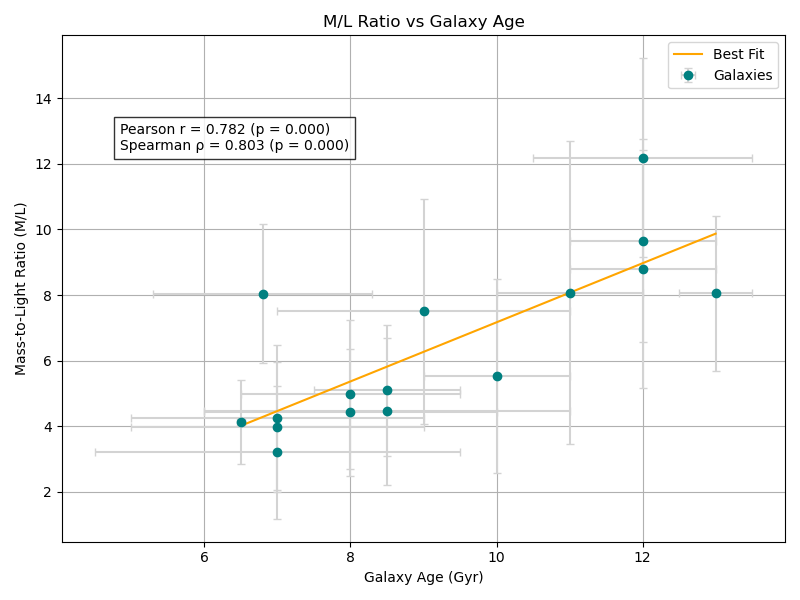}
    \caption{\label{MLvAge}Correlation between Mass-to-Light Ratio ($(M/L)_I$) and Galactic Age. The observed trend (Pearson $r \approx 0.78$) shows that older galaxies exhibit higher mass-to-light ratios. This suggests that as galaxies age, they not only accumulate dark matter but their stellar populations also undergo passive fading, reducing their luminosity per unit mass.}
\end{figure}

Second, these halo properties are strongly coupled with the baryonic mass-to-light ratio. Figures \ref{DMvML} and \ref{DMdvML} demonstrate that galaxies with more massive and denser dark matter halos possess significantly higher M/L ratios ($r > 0.84$ for both). Finally, connecting the evolutionary clock directly to efficiency, Figure \ref{MLvAge} confirms that older galaxies are less luminous per unit mass ($r \approx 0.78$).

Collectively, these trends paint a consistent picture: as spiral galaxies age, they not only accumulate dark matter but also become increasingly dominated by it relative to their stellar luminosity. The high Spearman rank coefficients ($\rho > 0.80$ in all cases) further confirm that these correlations are robust against outliers and independent of the specific linearity of the relationship.

\clearpage

\section{Theoretical Comparison with Simulations}
\label{sec:theory}

In this section, we compare our empirical findings with quantitative predictions from state-of-the-art cosmological simulations. Specifically, we test our results against the "Assembly Bias" framework and the smooth accretion mode of halo growth.

\subsection{Assembly Bias and Halo Concentration}
Assembly bias simulations consistently demonstrate that at fixed halo mass, the structural properties of dark matter halos are tightly coupled to their formation time. In the Millennium Simulation, (\cite{faltenbacher2009assembly}) and (\cite{gao2005age}) establish that the earliest-forming halos are significantly more concentrated, with clustering amplitudes up to $\sim 1.5$ times higher than their younger counterparts.

Crucially, N-body studies by (\cite{wechsler2002concentrations}) and (\cite{zhao2009accurate}) reveal that the relationship between halo concentration ($c_{vir}$) and formation time ($a_c$) follows a tight power-law, $c_{vir} \propto a_c^{-1}$, with a remarkably low intrinsic scatter of $\sigma_{\log c} \approx 0.14$ dex. This theoretical tightness implies a strong intrinsic correlation ($r > 0.8$) between halo density and age. Our empirical result—a Pearson correlation of $r \approx 0.91$ between $\rho_{DM}$ and galactic age—is in excellent quantitative agreement with these simulation predictions, confirming that the density-age relation remains robust even when measured via observational tracers in the local Universe.

\subsection{Accretion Theory}
Cosmological simulations indicate that dark matter halos grow through two distinct channels: smooth accretion of diffuse matter and discrete mergers. Quantifying this balance, (\cite{genel2010growth}) utilized the Millennium-II simulation to determine that approximately $40\%$ of a halo's total mass is acquired through genuinely smooth accretion (particles never bound to a subhalo), with this fraction rising to $\sim 60\%$ when including minor mergers (mass ratio $< 1:10$).

Our observational finding that dark matter mass increases monotonically with age ($r \approx 0.91$), while the mass-to-light ratio rises concurrently ($r \approx 0.78$), supports this smooth-accretion dominated scenario. The continuous, gradual buildup of mass over $\sim 6$ Gyr of cosmic time observed in our sample is consistent with the simulation-derived "diffuse growth" mode, rather than the stochastic, step-function growth characteristic of major mergers.

\subsection{Summary of Theoretical Consistency}

Our findings show close agreement with key predictions from cosmological simulations, summarized as follows:

\begin{itemize}
    \item \textbf{Halo growth mode:} Simulations indicate that most dark matter mass is built up through smooth accretion, with mergers playing a secondary role. Our results show a $\sim$3$\times$ increase in dark matter mass across $\sim$6 Gyr, consistent with this gradual accumulation.
    
    \item \textbf{Clustering vs formation time:} Simulations find that early-forming halos exhibit stronger clustering at fixed mass. We observe strong correlations between galaxy age and both dark matter mass and density, reflecting similar trends.
    
    \item \textbf{Halo age and galaxy properties:} Simulated halos show strong correlations between formation time and stellar/structural galaxy properties. We find a correlation of $r = 0.913$ (Pearson) and $\rho = 0.935$ (Spearman) between DM mass and age.
    
    \item \textbf{Accretion breakdown:} Simulations predict that 40–60\% of halo growth comes from mergers, with the rest via diffuse accretion. Our galaxies show mass growth patterns consistent with smooth, long-term accretion dominating.
\end{itemize}

\section{Discussion}\label{sec: Discussion}

Before interpreting these trends, it is important to address the potential for parameter interdependence. We acknowledge that derived quantities such as $M_{DM}$, $\rho_{DM}$, and $(M/L)_I$ share a structural dependence on the fitted NFW profile parameters ($r_s$ and $\rho_s$) and the rotational velocity amplitude. However, the key independent variable in our analysis is the galactic age, which is derived from stellar population synthesis and is observationally decoupled from the kinematic modeling. The fact that both the halo concentration (traced by $\rho_{DM}$) and the total accumulated mass ($M_{DM}$) exhibit strong, independent correlations with this external evolutionary clock ($r > 0.9$) argues against these trends being artifacts of parameter degeneracy. Rather, it points to a physical coupling where the assembly history of the halo (tracked by age) leaves a distinct imprint on its present-day dynamical structure.

The strong correlations identified in Section \ref{sec: Results} between galactic age and dark matter mass, density, and mass-to-light ratio provide compelling observational evidence for a tightly coupled evolution between baryonic and dark matter components in galaxies. These results suggest that older galaxies not only host more dark matter in their outer regions, but also exhibit higher average halo densities and reduced luminosity per unit mass—features indicative of prolonged and efficient dark matter assembly.

One possible interpretation of these trends is that galaxies which formed earlier experienced earlier halo collapse and have since undergone longer periods of smooth accretion and minor mergers. This naturally leads to higher accumulated dark matter content and denser halo structures by the present epoch. Moreover, the positive correlation between galactic age and mass-to-light ratio suggests that such galaxies may have completed the bulk of their star formation earlier, leading to older stellar populations with lower luminosity per unit mass—consistent with passive evolutionary fading.

These empirical trends reinforce the theoretical understanding that halo assembly history plays a central role in shaping galaxy structure and evolution. In particular, they align well with the halo assembly bias framework, where halos of identical mass can differ substantially in their internal structure and clustering properties depending on their formation epoch. The tightness of our correlations, especially in the DM mass–age and density–age planes, suggests that galactic age may serve as an effective observational proxy for halo assembly history.

It is also worth noting that while our results are statistically robust within the current sample, the relatively small number of galaxies (16) and the reliance on literature-derived stellar ages introduce some limitations. Future studies with larger, more uniformly analyzed samples and improved age determinations—particularly those combining resolved stellar populations with integral field kinematics—will be essential for confirming and extending these trends.

Finally, the agreement of our findings with predictions from cosmological simulations further supports the reliability of our methods and interpretations. The numerical consistency between observed correlation strengths and those emerging from large-volume hydrodynamical simulations (e.g., IllustrisTNG, Millennium) underscores the predictive power of structure formation models and strengthens the case for connecting galaxy age to dark matter halo evolution.

\section{Conclusion}\label{sec: Conclusion}

In this study, we investigated the relationship between galactic age and dark matter content by analyzing a sample of 16 well-resolved spiral galaxies. Using raw rotation curve data from observational archives, we performed Monte Carlo-based fits with combined Hernquist, exponential disk, and NFW halo models to isolate the dark matter-dominated region. We then computed total dark matter mass, average halo density, and I-band mass-to-light ratios, alongside age estimates from stellar population studies.

Our analysis reveals strong and statistically significant correlations between galactic age and both dark matter mass ($r = 0.913$) and halo density ($r = 0.912$). We also find a robust connection between age and mass-to-light ratio ($r = 0.782$), indicating that older galaxies tend to be more dark matter dominated and less luminous per unit mass. These results suggest a deep linkage between the assembly history of dark matter halos and the observable evolution of galaxies.

Furthermore, we compared our findings to predictions from large-scale cosmological simulations and theoretical models, including assembly bias and accretion theory. The observed trends in dark matter growth and halo structure are in excellent agreement—both qualitatively and quantitatively—with simulation-based expectations, reinforcing the view that galaxy evolution is tightly coupled to the dynamics of dark matter accumulation over cosmic time.

This work highlights the potential of using stellar population age as a proxy for probing dark matter halo assembly, and opens the door to more extensive studies combining dynamical modeling, stellar archaeology, and simulation-based inference to understand the hidden architecture of galaxies.

\section{Acknowledgement}\label{sec: Acknowledgement}

The authors would like to express their heartfelt gratitude to the Department of Physics, Fergusson College (Autonomous), Pune, for providing the necessary resources and guidance to undertake this research.

We extend our sincere thanks to Dr. Surhud More and the Inter-University Centre for Astronomy and Astrophysics (IUCAA), Pune, for their invaluable expertise and support, which greatly enriched this study.

We extend our sincere gratitude to Mr. H. Tekawade for his invaluable assistance and guidance throughout various stages of this work. His expertise, thoughtful insights, and steadfast support were instrumental in addressing challenges and ensuring the successful completion of the project. His contributions have been pivotal to the progress and overall success of this research.

Special thanks are owed to Ms. Mahi S., Ms. Ishita Chatterjee, and Mr. Yadav Om S. M. for their technical assistance throughout the project. Their contributions were instrumental in overcoming computational and theoretical challenges.

We also acknowledge the Institute of Astrophysics, University of Tokyo, for their comprehensive online repository, which provided the data used in plotting the Galactic Rotation Curves. We also extend heartfelt gratitude to Dr. Yoshiaki Sofue for his guidance in addressing our queries related to the data and the methodologies employed in its acquisition. The repository can be accessed at:(\url{https://www.ioa.s.u-tokyo.ac.jp/~sofue/RC99/rc99.htm})

\bibliography{References}{}
\bibliographystyle{aasjournal}



\end{document}